\documentclass{PoS}
\usepackage{graphicx}

\def\ttl#1{{\it #1}}
\def\ttl#1{}

\def\geqx{\,\raisebox{-1.0ex}{$\stackrel{\textstyle >}{\sim}$}\,}

\newcommand{\bee}{\begin{equation}}
\newcommand{\ee}{\end{equation}}
\newcommand{\beea}{\begin{eqnarray}}
\newcommand{\eea}{\end{eqnarray}}

\def\g{\gamma}

\def\psibar{\overline\psi}
\def\tbeta{\tilde\beta}

\title{Gauge theories with fermions in two-index representations}

\ShortTitle{Gauge theories with fermions in two-index representations}

\author{Thomas DeGrand\\
Department of Physics, University of Colorado, Boulder, CO 80309, USA\\
E-mail: \email{degrand@pizero.colorado.edu}}

\author{\speaker{Yigal Shamir} and Benjamin Svetitsky\\
        Raymond and Beverly Sackler School of Physics and Astronomy\\
        Tel-Aviv University, Ramat~Aviv, 69978~Israel\\
        E-mail: \email{shamir@post.tau.ac.il}, \email{bqs@julian.tau.ac.il}}

\abstract{
  After some introductory comments on the peculiar features of
  slowly running theories, I will report results obtained using the
  Schr\"odinger functional technique for two gauge theories that are
  believed to lie near the bottom of the conformal window: the SU(3)
  theory with two adjoint Dirac fermions, and the SU(4) theory with six
  Dirac fermions in the two-index antisymmetric representation. In both
  cases we find a small beta function in strong coupling, but we cannot
  confirm or rule out an infrared fixed point. In both theories the mass
  anomalous dimension levels off, staying well below 0.5, much like the
  theories with fermions in the two-index symmetric representation
  investigated earlier.}

\FullConference{31st International Symposium on Lattice Field Theory
  LATTICE 2013\\
  July 29 -- August 3, 2013\\
  Mainz, Germany}

\begin{document}


\section{Introduction}
Asymptotically free gauge theories can differ from QCD in several ways:
the number of colors $N_c$, the number of flavors $N_f$,
and the fermions' representation.  These theories are interesting for purely
theoretical reasons, and also as templates for physics
beyond the Standard Model.
There is a growing body of numerical work devoted to them (for a recent
review, see Ref.~\cite{JK}).

As the number of flavors is increased, typically the two-loop
coefficient of the beta function
\bee
\beta(g^2)= -b_1\frac{g^4}{16\pi^2}-b_2\frac{g^6}{(16\pi^2)^2} + \cdots,
\label{ptbeta}
\ee
will flip sign before the one-loop coefficient.
The range of $N_f$-values where $b_1>0>b_2$ defines perturbatively the
conformal window, where the running coupling is driven to an
infrared-attractive fixed point (IRFP).  Unlike QCD, in that case
no physical scale is generated dynamically, and the long-distance
behavior of all correlation functions is predicted to follow a power law.

The existence of an IRFP requires nonperturbative confirmation.
Particularly interesting are borderline theories in which $N_f$
is close to the critical value where the perturbative conformal window
is entered.

We have carried out a long-term program of studying gauge theories
with varying number of colors, and with fermions in various two-index
representations.  We concentrate on two observables.
The first is the nonperturbative beta function,
which we define and measure through the Schr\"odinger functional (SF) scheme.
The second is the mass anomalous dimension $\g_m$, which we define as usual
from the scaling behavior of $\psibar\psi$.  Thanks to chiral symmetry
(of the massless continuum theory), we may in fact extract $\g_m$
from the scaling of the isospin-triplet pseudoscalar density,
which in turn is much better
behaved on the lattice, and which we have measured on the same ensembles
used to determine the running coupling.

Previously, we studied gauge theories with fermions in the symmetric
two-index representation
\cite{DeGrand:2011qd,DeGrand:2012qa,DeGrand:2012yq,talks}.
Here we will report on two more theories with fermions
in a two-index representation \cite{DeGrand:2013uha}.  These are
the SU(3) theory with $N_f=2$ Dirac fermions in the adjoint representation,
and the SU(4) theory with $N_f=6$ Dirac fermions in the antisymmetric
representation, which is a sextet.  Choosing $N_f=6$ places that theory
near the bottom of the perturbative conformal window.

\section{Slow running}

The SF setup was originally developed aiming for a precise
numerical determination of the evolution of the QCD coupling.
Using the SF setup in a different gauge theory is straightforward.
But our analysis tools must be adapted to a new
situation where the coupling hardly runs at all.

\begin{figure}[t]
\begin{center}
\includegraphics[width=.4\linewidth]{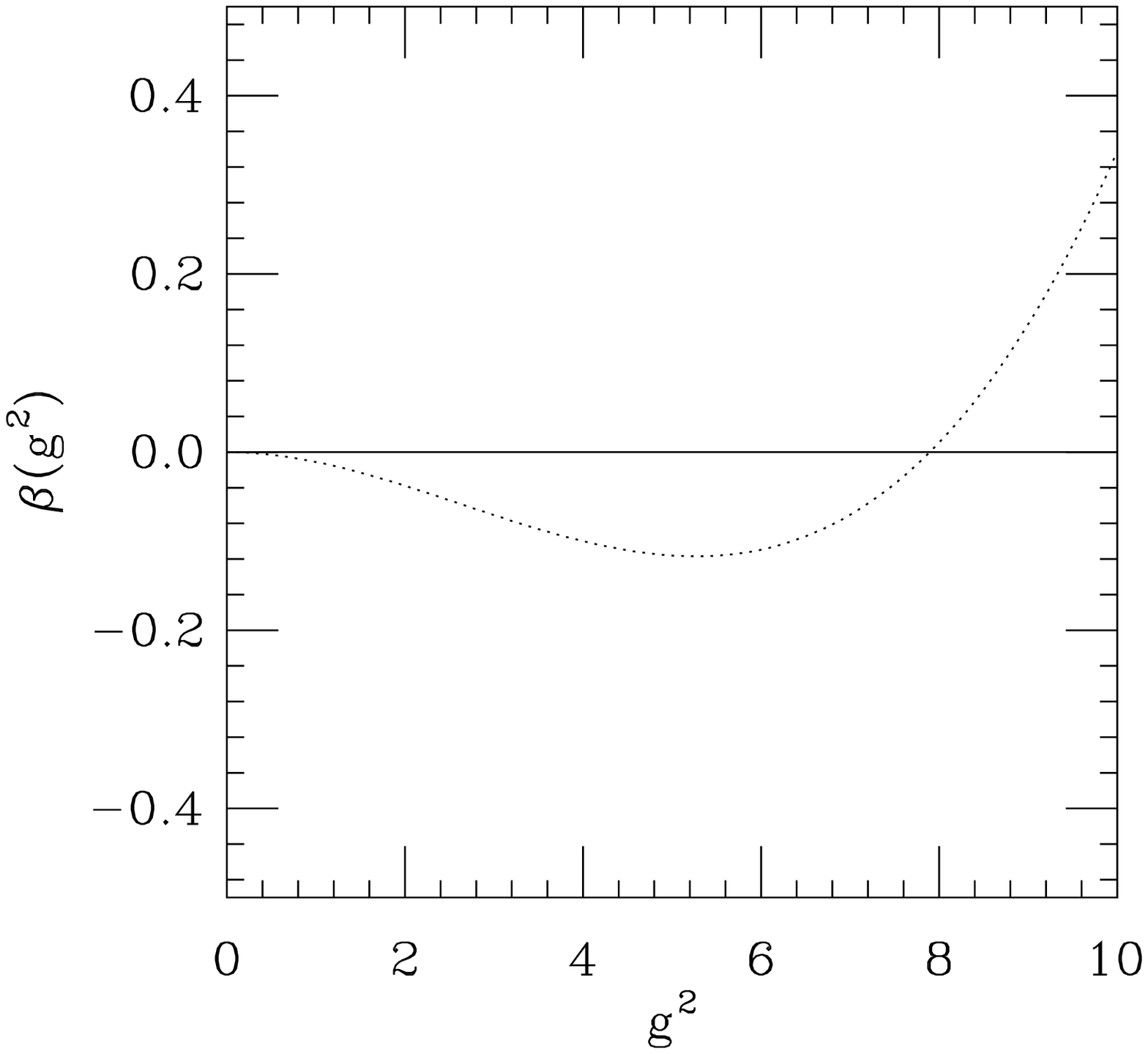}
\hspace{0.1\linewidth}
\includegraphics[width=.38\linewidth]{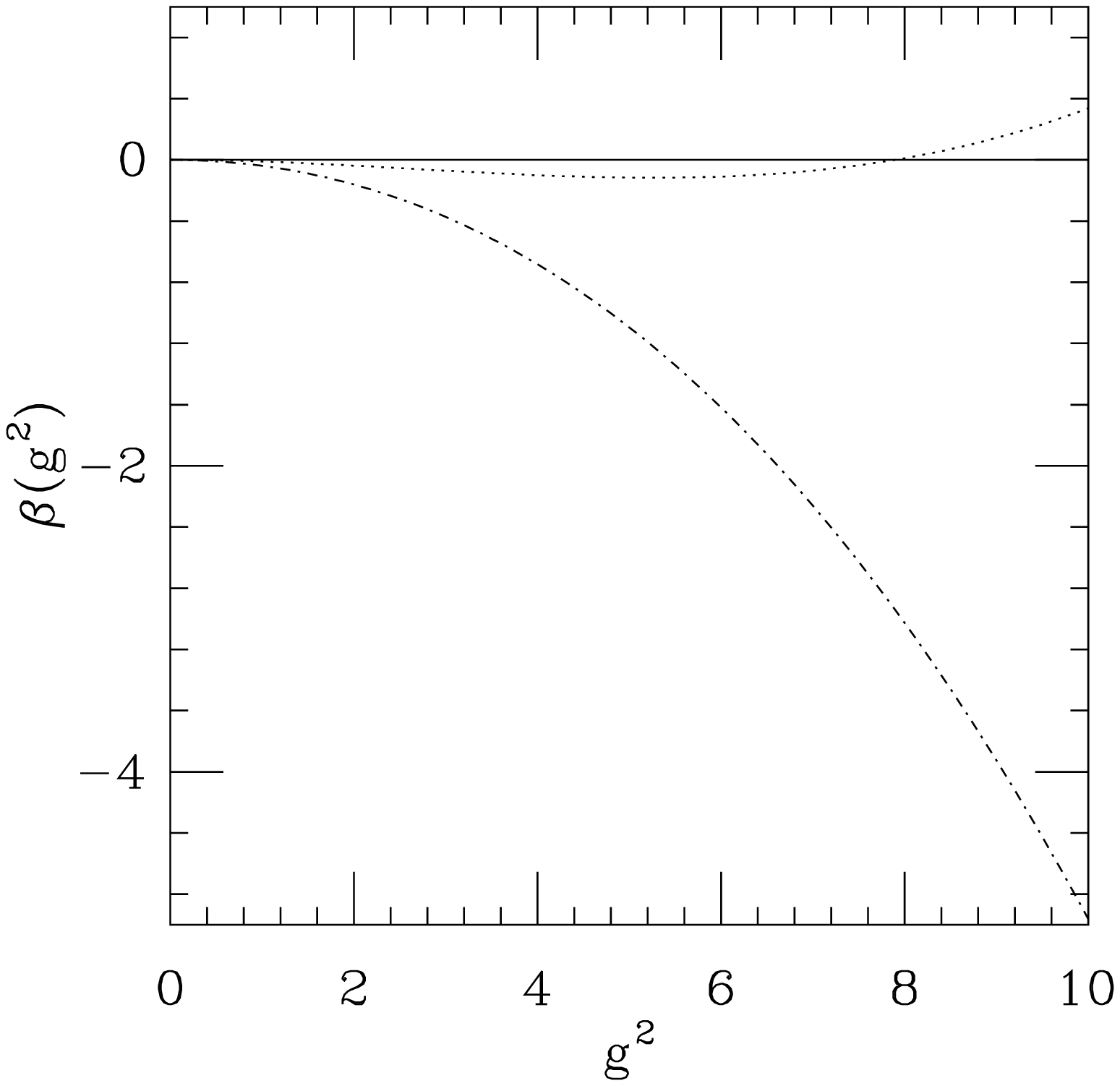}
\end{center}
\vspace*{-1ex}
\caption{Two-loop beta function in SU(2) gauge theories.
See text for explanation.}
\label{fig:bPT}
\end{figure}

To appreciate this difference consider Fig.~\ref{fig:bPT}, where,
as an example,
we show the two-loop beta function for two different SU(2) theories,
each containing two Dirac fermions in a given representation.
In the left panel the fermions are in the adjoint representation.
The (perturbative) fixed point is clearly visible.
In the right panel, the downward pointing curve is the beta
function of the theory with fermions in the fundamental representation.
Much like QCD, this beta function is always negative,
and grows in absolute value with increasing coupling.

The other curve in the right panel, which embraces the horizontal axis,
is once again the beta function of the adjoint-fermions theory.
The visual difference relative to the left panel comes from
the different vertical scales.  The lesson from this comparison
is that measuring a beta function so much smaller than
that of a QCD-like theory is bound to be substantially more difficult.

The prime dynamical question about any massless asymptotically free theory
is whether its infrared physics is conformal, or, alternatively, confining
and with broken chiral symmetry.
When the coupling runs slowly, in order to probe
interesting values of the renormalized coupling already the bare coupling
must be quite strong.  As a result, unlike in QCD simulations,
lattice perturbation theory is not applicable at the lattice scale
in our simulations, and cannot provide us with any guidance.
At the same time, as we will see, new analysis methods can be developed
that are especially tailored to slow running.

\section{Nonperturbative beta function}

We use Wilson-clover fermions.  The links in the Dirac operator
are nHYP smeared links that are subsequently promoted
to the fermions' representation.  The geometry of our SF lattices
is hypercubical with equal size $L=Na$ in all four directions.
For most values of the bare parameters studied, we performed
simulations for $N=6,8,10,12,16$.
Full details can be found in Ref.~\cite{DeGrand:2013uha}.

\begin{figure}[t]
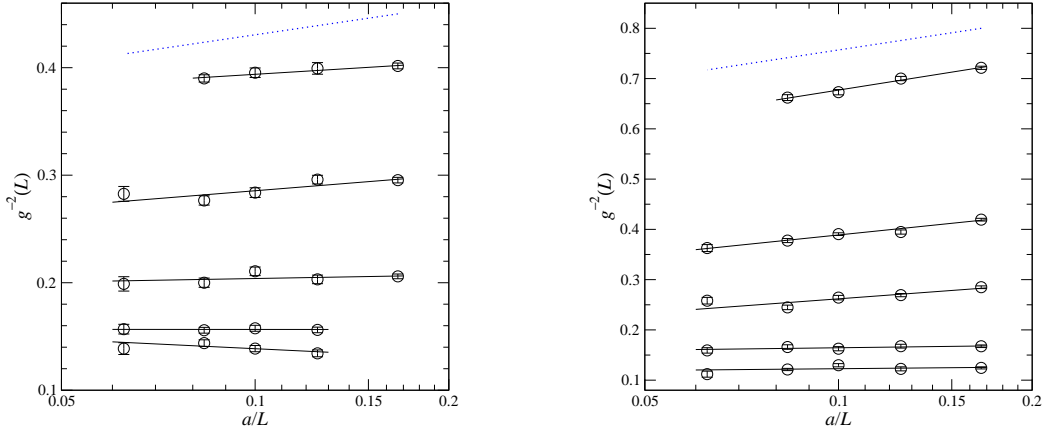

\vspace*{1ex}
\begin{center}
\includegraphics[width=.4\linewidth]{SU3_1g2_vs_lcombined.eps}
\hspace{0.1\linewidth}
\includegraphics[width=.4\linewidth]{SU4_1g2_vs_lcombined.eps}
\end{center}
\vspace*{-1ex}
\caption{
A semilog plot of the running coupling $1/g^2\;$ vs. $\,a/L$
in the SU(3)/adjoint theory (left) and the SU(4)/sextet theory (right).}
\label{fig:1g2}
\end{figure}

Instead of the usual beta function~(\ref{ptbeta}), it is convenient
to introduce the beta function $\tbeta(u)$ for $u\equiv1/g^2$, define as
\bee
  \tbeta(u) \equiv \frac{d(1/g^2)}{d\log L}
  = 2\beta(g^2)/g^4 = 2u^2 \beta(1/u).
\label{invbeta}
\ee
Were the beta function $\tbeta(u)$ constant, the running coupling would take
the form
\bee
u(L)=c_0+c_1 \log \big(L/(8a)\big)\ ,
\label{linfit}
\ee
where $c_0$ is $u(L=8a)$, and $c_1$ is the constant value of $\tbeta(u)$.
In Fig.~\ref{fig:1g2} we show our results for the running coupling
in the two theories.  The straight lines are fits to Eq.~(\ref{linfit})
of the results from all volumes at each fixed set of bare parameters.
It is evident from this figure that, over the range
of volumes we studied, a constant beta function is a reasonable
first approximation of the data.  As we go upwards in the figure,
both the bare and the renormalized couplings get smaller.
For reference, the dotted blue lines
show the slopes of one-loop running (note that $\tbeta(u)$ is
constant in the one-loop approximation).

In order to extrapolate our results to the continuum limit we make
use of the following observation.  If the coupling did not run at all,
the only obstruction to Eq.~(\ref{linfit}) would be discretization
errors.  Much like in a free theory, in the absence of any dynamical scale
the discretization
errors would necessarily depend on $a/L$ only.  Indeed we may then
identify the lattice spacing $a$ with $1/L_{min}$,
where $L_{min}$ is the smallest
lattice size included in the fit~(\ref{linfit}).  By repeatedly
dropping the smallest lattice, we should get better and better estimates
of the continuum-limit value.  Ordering all lattice sizes as
$L_1<L_2<\ldots<L_n$, we denote by $c_0^{(k)},c_1^{(k)}$ the parameters
obtained from a fit in which the smallest size kept was $L_k$.
We can then extrapolate to $a/L=0$ either linearly,
\begin{equation}
c_1^{(k)}=\tbeta(u)+C (a/L_k)
\label{linex}
\end{equation}
or quadratically,
\begin{equation}
c_1^{(k)}=\tbeta(u)+C (a/L_k)^2.
\label{quadex}
\end{equation}

\begin{figure}[t]
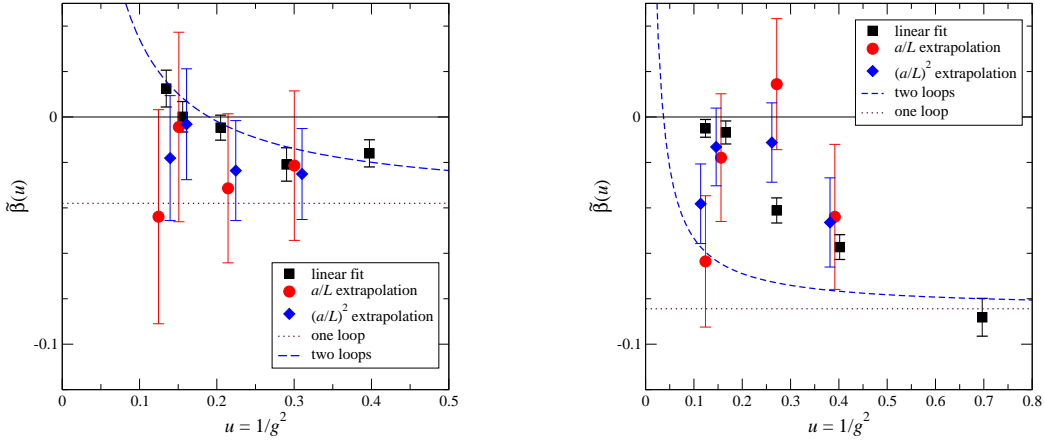

\vspace*{1ex}
\begin{center}
\includegraphics[width=.4\linewidth]{SU3adjbeta_scaled.eps}
\hspace{0.1\linewidth}
\includegraphics[width=.4\linewidth]{SU4as2beta.eps}
\end{center}
\vspace*{-1ex}
\caption{Beta function $\tbeta(u)$ of the SU(3)/adjoint theory (left)
and the SU(4)/sextet theory (right), extrapolated to the continuum limit.
The points for the extrapolations have been displaced slightly for clarity.}
\label{fig:beta}
\end{figure}

The results of both types of extrapolation, along with the results
of the simple fit~(\ref{linfit}), are shown in Fig.~\ref{fig:beta}.
As can be seen, substantially bigger computation resources and/or
better observables would be required to establish the presence or absence
of an IRFP in these theories.

For a very small lattice spacing $a$, or, equivalently, for very large $L$,
ultimately the linear discretization error must dominate.
As it turns out, even in the one-loop approximation linear and quadratic
discretization errors remain comparable in size over the entire range
of volumes we have.  (We discussed this in some detail regarding a
different slowly-running theory in Ref.~\cite{DeGrand:2012yq}.)
Therefore there is no good reason to prefer one type of error over the other,
and, in principle, we must allow for linear and quadratic discretization
errors simultaneously.  Since our data are not precise enough to allow
for such a combined extrapolation, the results of both
types of extrapolation must be considered as models.

Equation~(\ref{linfit}) is exact only in the limit of
a constant beta function $\tbeta(u)$.  In reality, as we have
discussed in Ref.~\cite{DeGrand:2011qd}, the slow change
in $\tbeta(u)$ will give rise to higher powers of $\log L$.
As a better approximation of the continuum evolution we may therefore take
\bee
u(L)=c_0 + c_1 \log \big(L/(8a)\big) + c_2 \log^2 \big(L/(8a)\big)\ .
\label{quadfit}
\ee
In principle we could then perform a similar continuum-extrapolation
procedure by repeatedly dropping the smallest volumes, now using
Eq.~(\ref{quadfit}) as our basic fit.
Once again, our data are not precise enough
to obtain meaningful results this way.  We stress, however, that
the extrapolations~(\ref{linex}) and~(\ref{quadex}) based on
Eq.~(\ref{linfit}) both have good quality, showing that a term like $\log^2$
is unnecessary given our statistical error.

\section{Mass anomalous dimension}

\begin{figure}[t]
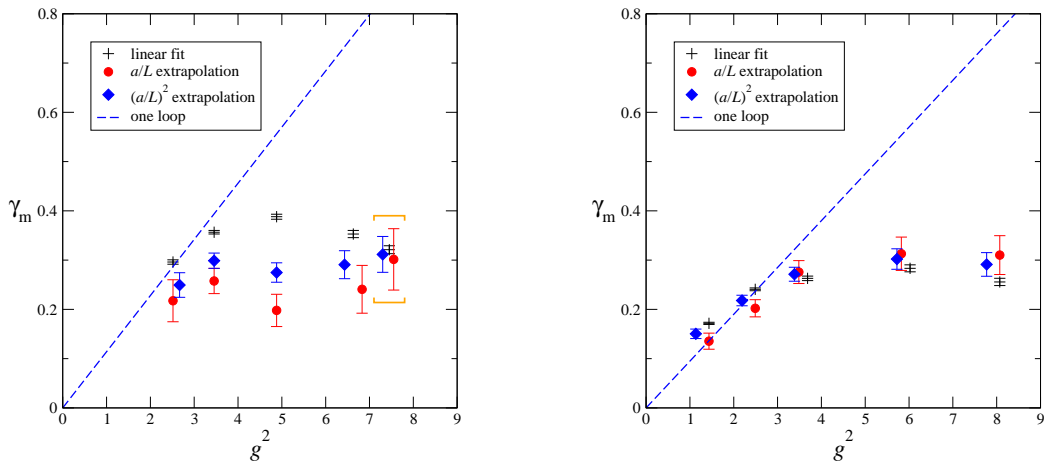

\vspace*{1ex}
\begin{center}
\includegraphics[width=.4\linewidth]{SU3adjgamma.eps}
\hspace{0.1\linewidth}
\includegraphics[width=.4\linewidth]{SU4as2gamma.eps}
\end{center}
\vspace*{-1ex}
\caption{Mass anomalous dimension $\gamma_m(g^2)$ of the SU(3)/adjoint theory
(left) and the SU(4)/sextet theory (right),
plotted as a function of $g^2(L=8a)$.}
\label{fig:gamma}
\end{figure}

In the approximation that the coupling does not run at all,
the pseudoscalar renormalization constant $Z_P$ follows a power law.
Accordingly, for each set of bare parameters, we fit
\bee
\log Z_P(L)=c_0 + c_1 \log \big((8a)/L\big)\ ,
\label{linfitZ}
\ee
where now $c_1$ gives an estimate for the mass anomalous dimension $\g_m$.
We plot the results of these fits in Fig.~\ref{fig:gamma}, together
with the results of linear and quadratic continuum extrapolations following
the same procedure as before.  Unlike the beta function, here the error
bars remain quite small even after the continuum extrapolation.

Focusing first on the SU4/sextet theory, we see that at weak coupling
our results agree with one-loop perturbation theory.  But for $g^2\geqx 3$,
$\g_m$ levels off, becoming practically independent of $g^2$.
A similar behavior, although a bit noisier, is seen in the SU(3)/adjoint
theory.  [In the case of the rightmost (strongest coupling) point,
we could not overcome the long autocorrelations of the observable.
The results marked by the orange brackets come from 3 streams that agreed
with each other, after discarding an outlier stream \cite{DeGrand:2013uha}.]

The leveling off of $\g_m$ is a remarkable feature, common to all of
the theories with fermions in two-index representation we have studied
\cite{DeGrand:2011qd,DeGrand:2012qa,DeGrand:2012yq,DeGrand:2013uha}.
This is a surprising result, that, to our knowledge, was not predicted
by any perturbative calculation.

\section{$N_c$ scaling}

In the course of our work we have studied two gauge theories each
containing two Dirac fermions in the adjoint representation:
the SU(2) theory \cite{DeGrand:2011qd} and, here, the SU(3) theory.
It is interesting to look for trends as $N_c$ is changed \cite{Shifman:2013yca}.

\begin{figure}[t]
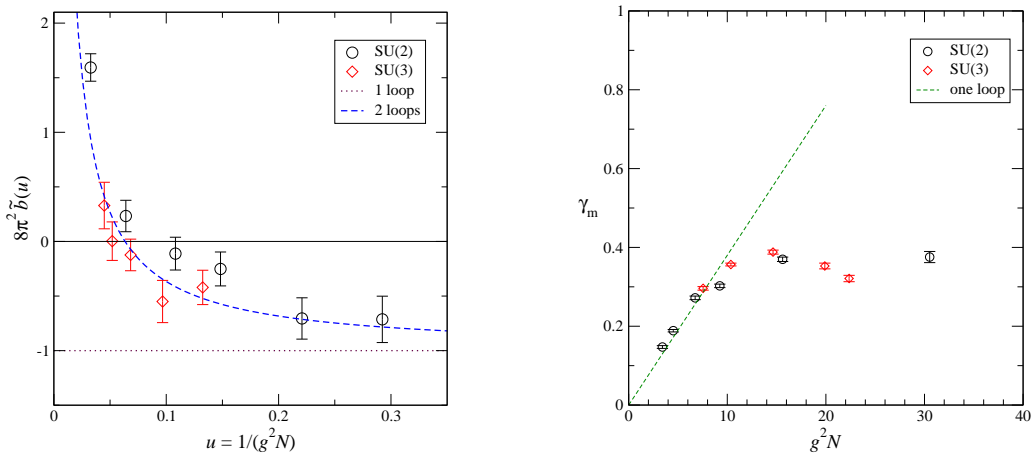

\vspace*{1ex}
\begin{center}
\includegraphics[width=.39\linewidth]{adjbetalargeN.eps}
\hspace{0.1\linewidth}
\includegraphics[width=.4\linewidth]{adjgammalargeN.eps}
\end{center}
\vspace*{-2ex}
\caption{$N_c$ scaling.}
\label{fig:Ncscale}
\end{figure}

Fig.~\ref{fig:Ncscale} shows this comparison.  Here we only
compare the basic linear fits~(\ref{linfit}) for the beta function
and~(\ref{linfitZ}) for the mass anomalous dimension.
The results suggest that large-$N_c$ scaling works quite well down
to the smallest value $N_c=2$ (including
any discretization error that is present in the plots).
We note, however, that unlike the SU(2) theory, where we established
the existence of an IRFP,
the SU(3) theory could be confining \cite{Karsch:1998qj}.

\section{Conclusions}

While somewhat disappointing, in view of the difficulties explained in Sec.~2
it is no surprise that the extrapolations of our data for
the nonperturbative beta functions result in rather large errors.

Our results for the mass anomalous dimension are much nicer.
They have fairly small errors even after the continuum extrapolation.
The surprising leveling off at strong coupling leads to
a scheme-independent universal bound $\g_m < 0.5$,
a bound that applies to {\it all}
the theories we have studied in the course of this research program.

A second look at the continuum extrapolations of the beta function
of the SU(4)/sextet theory (Fig.~\ref{fig:beta}, right panel) may reveal
a hint of the behavior known as ``walking,''
where the beta function first gets very close to zero, and then veers off.
Accordingly, after many decades of almost no running, eventually
the couplings grows strong enough to trigger
chiral symmetry breaking and confinement.
Walking theories can naturally accommodate a light composite scalar,
which can arise as a pseudo Nambu--Goldstone boson of
the spontaneously broken approximate dilatation symmetry.
For a recent discussion of whether this scalar
could be identified with the 125 GeV particle discovered at the LHC,
see Ref.~\cite{JK}.

Even if walking technicolor can explain the existence of a light Higgs
particle, to qualify as a successful theory of Electro-Weak symmetry
breaking it has to provide a mechanism
for the generation of lepton and quark masses as well.  Traditionally,
this was done by invoking an ``extended'' technicolor theory.
For this mechanism to meet phenomenological constraints,
typically a large mass anomalous dimension, $\g_m \approx 1$,
was invoked.  Our results for $\g_m$ therefore cast doubt on the ability
to use any of the theories we have studied as (extended) technicolor
candidates.

\acknowledgments
This work was supported in part by the Israel Science Foundation
under grant no.~423/09 and by the U.~S. Department of Energy
under grant DE-FG02-04ER41290.  We refer to \cite{DeGrand:2013uha}
for acknowledgments of the supercomputer resources we have used.

\end{document}